\documentclass[prd,singlecolumn,amsmath,amssymb,floatfix,showpacs,notitlepage,nofootinbib,preprintnumbers]{revtex4-1}

\usepackage[T1]{fontenc}
\usepackage[utf8]{inputenc}
\usepackage{bm}
\usepackage{hyperref}
\usepackage{graphicx}
\usepackage{color}

\newcommand{\sfH}{\mathsf{H}} 
\newcommand{\bnabla}{\boldsymbol{\nabla}}
\newcommand{\thmax}{\vartheta_\mathrm{max}}
\newcommand{\kmax}{\kappa^\mathrm{max}}
\newcommand{\p}[1]{\protect{(#1)}}

\begin{document}

\title{Flavor instabilities in the multi-angle neutrino line model}

\author{Sajad Abbar}
\email{sabbar@unm.edu}
\author{Huaiyu Duan
}
\email{duan@unm.edu}
\author{Shashank Shalgar}
\email{shashankshalgar@unm.edu}
\affiliation{Department of Physics \& Astronomy, University of New
  Mexico, Albuquerque, NM 87131, USA}

\date{\today}

\begin{abstract}
Neutrino flavor oscillations in the
presence of ambient neutrinos is nonlinear in nature which leads
to interesting phenomenology that has not been well understood. 
It was recently shown that,  in the
two-dimensional, two-beam neutrino Line model, the inhomogeneous
neutrino oscillation modes on 
small distance scales can become unstable at larger neutrino densities than the
homogeneous 
mode does. We develop a numerical code to solve neutrino
oscillations in the multi-angle/beam Line model with a continuous
neutrino angular
distribution. We show that the inhomogeneous oscillation modes can
occur at even higher neutrino densities in the multi-angle model than
in the two-beam model.  We also find that the inhomogeneous  modes on
sufficiently small scales can be unstable at smaller neutrino
densities with ambient matter than without, although a larger matter
density does shift the instability region of the homogeneous mode to 
 higher neutrino densities in the Line model
as it does in the one-dimensional supernova Bulb model. 
Our results suggest that the inhomogeneous neutrino oscillation modes can be
difficult to treat numerically because the problem of spurious
oscillations becomes more severe for oscillations  on smaller
scales.
\end{abstract}


\preprint{INT-PUB-15-039}
\maketitle

\section{Introduction}
The observation of neutrino flavor oscillations has established that
neutrinos have non-vanishing masses and that their propagation (or
mass) eigenstates are linear combinations of weak-interaction
states. One of the remarkable successes in the field of
experimental particle physics has been the measurement of all neutrino mixing
parameters except the sign of the atmospheric mass-squared difference
and the $\mathcal{CP}$ violation phase.  

Neutrino flavor oscillations can be modified by the potential due to
the presence of electrons
and nucleons in the medium \cite{Mikheev:1986gs,Wolfenstein:1977ue} or
the presence of ambient neutrinos
\cite{Fuller:1987aa,Notzold:1987ik,Pantaleone:1992eq}. There are 
two major differences in the phenomenology of neutrino flavor
oscillations due to the presence of ambient neutrinos as opposed to
ordinary matter. Firstly, unlike ordinary matter the presence of
ambient neutrinos makes the neutrino flavor evolution nonlinear in
nature. Secondly, in the case of ordinary matter the electrons and
nucleons are usually non-relativistic, and the potential experienced
by neutrinos is independent of direction to a very good approximation
(see~\cite{Mohapatra:1991ng} for a review). This is not true in the
case of neutrino-neutrino self-interaction. These
two differences make neutrino oscillations
in the neutrino medium very interesting and at the same time a
challenging problem solve. It has been shown that 
a dense neutrino gas  can undergo flavor oscillations collectively
\cite{Kostelecky:1993dm,Kostelecky:1993yt}. 
 
The effects of neutrino-neutrino interaction can be important in
extreme environments with large neutrino densities like that in the
interior of a core-collapse supernova.  
It was discovered in numerical simulations that the neutrino flavor
evolution can be 
dramatically different for the normal and inverted hierarchies inside
supernovae~\cite{Duan:2006jv, Duan:2006an}. 
However, in order to make the numerical simulations manageable, a
simplified one-dimensional supernova model called the (neutrino) Bulb model was
used. There are several effects that are not taken into account in the
Bulb model, although they can modify neutrino
oscillations significantly. For example, it has been found that the
neutrino emission with an (approximate) axial symmetry around the
radial direction can evolve into a configuration with a large axial
asymmetry~\cite{Raffelt:2013rqa,Mirizzi:2013rla}. It has also been
suggested that the back-scattering of neutrinos from the nucleons in
the envelope of the supernova can lead to significant modification of
the neutrino potential~\cite{Cherry:2012zw}.  

In this paper we investigate the physics of collective neutrino
oscillations in the neutrino Line model with two spatial
dimensions. The study of this model can provide us with useful 
insights into the qualitative differences of the phenomenology of
collective neutrino oscillations in models with one and multiple
spatial dimensions. This study is a generalization of the work done
for the two-beam Line model in
Ref.~\cite{Duan:2014gfa} where only two neutrino beams are emitted
from each neutrino source point.

\section{The neutrino Line model}
\label{sec:model}
\subsection{Equations of motion}
\begin{figure}
\begin{center}
\includegraphics[width=0.5\textwidth]{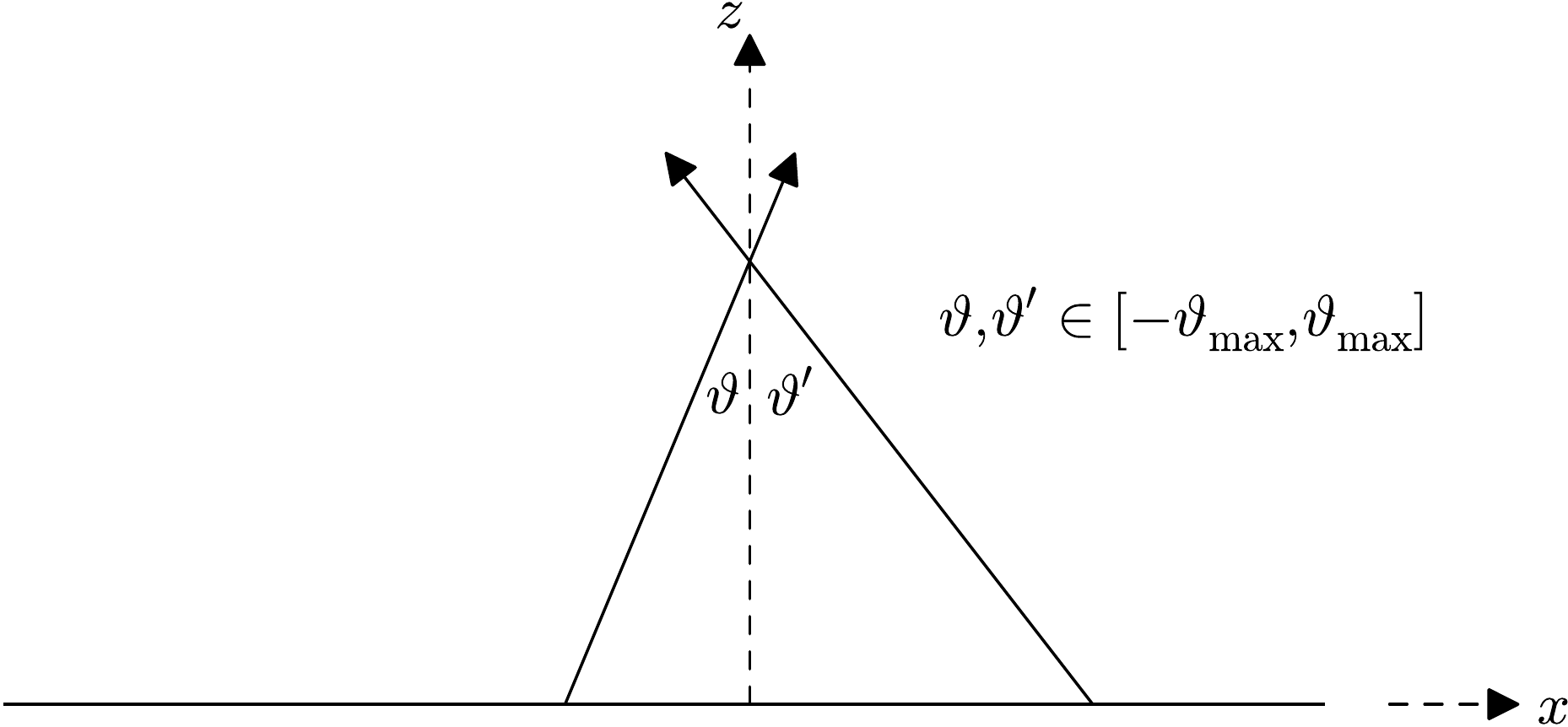}
\end{center}
\caption{A schematic diagram of the two dimensional (neutrino) Line
  model. Each point on the $x$-axis or the ``neutrino Line'' emits neutrino
  beams with emission angles $\vartheta$ within range
  $[-\vartheta_\text{max}, \vartheta_\text{max}]$.}
\label{geom}
\end{figure}
In the stationary, two-dimensional (neutrino) Line model neutrinos and
antineutrinos are emitted 
from the $x$-axis or the ``neutrino Line'' and propagate in the $x$-$z$
plane (see Fig.~\ref{geom}). 
We assume that
the neutrinos and antineutrinos are of single energy $E$ and the same
normalized angular distribution $g(\vartheta)$ such that the number
fluxes of the neutrino and antineutrino within angle range
$[\vartheta,\vartheta+ d\vartheta]$ are 
$n_\nu g(\vartheta) d\vartheta$ and $n_{\bar\nu} g(\vartheta)
d\vartheta$, respectively, where $\vartheta$ is the emission angle of
the neutrino 
beam, and $n_\nu$ and $n_{\bar\nu}$ are the (constant) total number
densities of the neutrino and antineutrino, respectively.
The flavor
quantum states of the neutrino and antineutrino of emission angle
$\vartheta$ and at position $(x,z)$ are given by density matrices
$\rho_{\vartheta}(x,z)$ and $\bar{\rho}_{\vartheta}(x,z)$, respectively
\cite{Sigl:1992fn}. We use the normalization condition
\begin{align}
  \mathrm{tr}\rho = \mathrm{tr}\bar\rho = 1
\end{align}
such that the diagonal elements of a density matrix give the
probabilities for the neutrino or antineutrino to be in the
corresponding weak-interaction states.
With these conventions the self-interaction potential
for $\rho_\vartheta(x,z)$ in the Line model can be written as
\begin{align}
\sfH_{\nu\nu,\vartheta}(x,z) = \mu
  \int
  \left[\rho_{\vartheta^{\prime}}(x,z)-\alpha
  \bar{\rho}_{\vartheta^{\prime}}(x,z)\right]
[1-\cos(\vartheta-\vartheta^{\prime})]
  g(\vartheta')~d\vartheta^{\prime},
\label{h:self}
\end{align}
where $\mu=\sqrt{2}G_{F}n_{\nu}$ with $G_F$ being the Fermi coupling
constant, and 
$\alpha=n_{\bar\nu}/n_\nu$.
In the Line model the strength of the neutrino
self-interaction $\mu$ is constant. In realistic astrophysical
environments such as core-collapse supernovae,
however, $\mu$ can decrease with increasing
distance from the neutrino source.

The flavor evolution of the neutrino or antineutrino is
governed by the equation of motion (EoM)
\begin{eqnarray}
i \mathbf{v}_\vartheta \cdot \bnabla \rho_\vartheta = 
[\sfH_\mathrm{vac}+\sfH_\mathrm{mat}+\sfH_{\nu\nu,\vartheta},\,\rho_\vartheta],
\label{eq:inl}
\end{eqnarray}
where $\mathbf{v}_\vartheta$ is the unit vector that denotes the propagation
direction of the neutrino with emission angle $\vartheta$, and 
$\sfH_{\mathrm{vac}}$ and $\sfH_{\mathrm{mat}}$ are
the standard vacuum mixing Hamiltonian and matter potential,
respectively.
In this work we assume the mixing between two active neutrino flavors
$\nu_e$ and $\nu_\tau$ with small vacuum mixing angle
$\theta_\mathrm{v}\ll1$. 
Therefore,
\begin{eqnarray}
\sfH_\omega=\sfH_{\mathrm{vac}}+\sfH_{\mathrm{mat}} \approx
\frac{(\lambda-\eta\omega)}{2}
\begin{pmatrix}
1 & 0 \cr
0 & -1  
\end{pmatrix}
= \frac{(\lambda-\eta\omega)}{2}\sigma_3,
\label{h:vac}
\end{eqnarray}
where $\lambda=\sqrt2 G_F n_e$ with $n_e$ being
the net electron number density,
$\eta$ is a parameter which takes a value of either $+1$ or $-1$ for
the normal neutrino mass hierarchy (NH, the mass-squared
difference $\Delta m^2>0$) or the inverted hierarchy (IH, $\Delta m^2 <0$), and
$\omega= |\Delta m^2|/2 E$ is the vacuum oscillation frequency of the
neutrino.
Eq.~\eqref{eq:inl} can also be written in a more explicit form:
\begin{align}
i (\cos\vartheta \partial_{z}  + \sin\vartheta
\partial_{x})\rho_{\vartheta}
&= \frac{(\lambda-\eta\omega)}{2}[\sigma_3,\rho_{\vartheta}] 
+ \mu
\int
[1-\cos(\vartheta-\vartheta^{\prime})]
\left[\rho_{\vartheta^{\prime}} -\alpha \bar{\rho}_{\vartheta^{\prime}} ,\rho_{\vartheta} \right] g(\vartheta')~d\vartheta^{\prime}.
\label{eqmot:2}
\end{align}
The EoM for $\bar\rho_\vartheta$ is the same as Eq.~\eqref{eqmot:2}
except with replacement $\omega\rightarrow-\omega$.

As in Ref.~\cite{Duan:2014gfa} we impose a periodic boundary condition
along the $x$ axis such that $\rho_\vartheta(x+L,z)=\rho_\vartheta(x,z)$
and $\bar\rho_\vartheta(x+L,z)=\bar\rho_\vartheta(x,z)$.
It is convenient to recast the $x$-dependence of the neutrino density matrix
in terms of Fourier moments:
\begin{align}
\rho_{m,\vartheta}(z) &= \frac{1}{L}\int_{0}^{L} e^{-i k_m x}
  \rho_{\vartheta}(x,z) dx, 
& \bar{\rho}_{m,\vartheta}(z) &=
  \frac{1}{L}\int_{0}^{L} e^{-i k_m x} \bar{\rho}_{\vartheta}(x,z) dx,
\label{mom:def}
\end{align}
where $k_m=2\pi m /L$.
It is straightforward to derive the EoM in the moment basis which are
\begin{subequations}
\label{eq:moments}
\begin{align}
i\cos\vartheta \partial_{z} \rho_{m,\vartheta}  &=   k_m 
\sin\vartheta \rho_{m,\vartheta}  + \frac{(\lambda-\eta\omega)}{2}
[\sigma_3, \rho_{m,\vartheta} ] 
\nonumber\\
&\quad
+   \mu \sum_{m^{\prime}}
\int
\left[\rho_{m^{\prime},\vartheta^{\prime}}  - \alpha
  \bar{\rho}_{m^{\prime},\vartheta^{\prime}} ,
  \rho_{m-m^{\prime},\vartheta}  \right]
     [1-\cos(\vartheta-\vartheta^{\prime})]
g(\vartheta')~d\vartheta^{\prime}, \\
i\cos\vartheta \partial_{z} \bar{\rho}_{m,\vartheta}  &=  k_m
 \sin\vartheta \bar{\rho}_{m,\vartheta}  +
\frac{(\lambda+\eta\omega)}{2}
[\sigma_3 , \bar{\rho}_{m,\vartheta}]
\nonumber\\ 
&\quad +   \mu \sum_{m^{\prime}}
\int
\left[\rho_{m^{\prime},\vartheta^{\prime}}  - \alpha
  \bar{\rho}_{m^{\prime},\vartheta^{\prime}} ,
  \bar{\rho}_{m-m^{\prime},\vartheta}  \right]
     [1-\cos(\vartheta-\vartheta^{\prime})]
g(\vartheta')~d\vartheta^{\prime}. 
\end{align}
\end{subequations}

\subsection{Collective modes in the linear regime}
\label{sec:lin}
We assume that the neutrinos and antineutrinos are emitted from the
Line source in the electron flavor only. In the regime where neutrino
oscillations are insignificant, the neutrino density matrices have the
form
\begin{align}
\rho_\vartheta(x,z) &\approx 
\begin{pmatrix}
1 & \epsilon_\vartheta \\
\epsilon_\vartheta^* & 0
\end{pmatrix}, &
\bar\rho_\vartheta(x,z) &\approx 
\begin{pmatrix}
1 & \bar\epsilon_\vartheta \\
\bar\epsilon_\vartheta^* & 0
\end{pmatrix}.
\end{align}
When there is a flavor instability, the off-diagonal elements of the
density matrices grow exponentially, which can result in collective
neutrino oscillations. In this section we apply the method of flavor
stability analysis to the multi-angle Line model which was first developed in
Ref.~\cite{Banerjee:2011fj}.

In the moment basis we have
\begin{align}
\rho_{m,\vartheta}(z)  &\approx  
\begin{pmatrix}
\delta_{0,m} & \epsilon_{m,\vartheta}  \cr
\epsilon^{*}_{-m,\vartheta}  & 0
\end{pmatrix}, &
\bar{\rho}_{m,\vartheta}(z)  &\approx 
\begin{pmatrix}
\delta_{0,m} & \bar{\epsilon}_{m,\vartheta}  \cr
\bar{\epsilon}^{*}_{-m,\vartheta}  & 0
\end{pmatrix}
\label{den:mat}
\end{align}
Keeping only the terms up to $\mathcal{O}(\epsilon)$ in
Eq.~\eqref{eq:moments} we obtain
\begin{subequations}
\label{off:term}
\begin{align}
  i\cos\vartheta \partial_{z} \epsilon_{m,\vartheta}  
&=
  [k_m  \sin\vartheta +\lambda - \omega\eta +
  (1-\alpha)\tilde\mu_\vartheta] 
  \epsilon_{m,\vartheta}  
- \mu \int
  [1-\cos(\vartheta-\vartheta^{\prime})](\epsilon_{m,\vartheta^{\prime}}
  -\alpha \bar{\epsilon}_{m,\vartheta^{\prime}} ) g(\vartheta')
  d\vartheta^{\prime},\\
i\cos\vartheta \partial_{z} \bar{\epsilon}_{m,\vartheta}  
&=
  [k_m  \sin\vartheta   +\lambda+ \omega\eta  +
  (1-\alpha)\tilde\mu_\vartheta]
  \bar{\epsilon}_{m,\vartheta}
- \mu \int
      [1-\cos(\vartheta-\vartheta^{\prime})]
      (\epsilon_{m,\vartheta^{\prime}}  - \alpha
      \bar{\epsilon}_{m,\vartheta'} ) g(\vartheta')
      d\vartheta^{\prime},
\end{align}
\end{subequations}
where
\begin{align}
  \tilde\mu_\vartheta =  \mu\int
  [1-\cos(\vartheta-\vartheta')] g(\vartheta')\,d\vartheta'
\end{align}
is the effective strength of neutrino self-interaction for the
neutrino beam with emission angle $\vartheta$.
As in the two-beam model, the flavor evolution of the neutrino fluxes
in different moments is decoupled in the linear regime, although the
evolution of the neutrino moments with different emission angles
$\vartheta$ are still coupled.

Assuming that the $m$th neutrino moment oscillates with collective
oscillation frequency $\Omega_m$, we can write
\begin{align}
\epsilon_{m,\vartheta}(z) &= Q_{m,\vartheta} e^{-i\Omega_m z}, 
& \bar{\epsilon}_{m,\vartheta}(z) &= \bar{Q}_{m,\vartheta} e^{-i\Omega_m z},
\label{off:par}
\end{align}
where $Q_{m,\vartheta}$ and $\bar{Q}_{m,\vartheta}$ are
$z$-independent.
Applying this ansatz to Eq.~\eqref{off:term} we obtain
\begin{subequations}
\label{lin:q}
\begin{align}
D_m(\omega,\vartheta) Q_{m,\vartheta} 
&=   (a_{m} - c_m \cos\vartheta - s_m \sin\vartheta)\mu,\\
D_m(-\omega,\vartheta) \bar{Q}_{m,\vartheta}
&=   (a_{m} - c_m \cos\vartheta - s_m \sin\vartheta)\mu
\end{align}
\end{subequations}
or
\begin{subequations}
\label{q:sol}
\begin{align}
  Q_{m,\vartheta} &=
  \frac{(a_{m} - c_m \cos\vartheta - s_m  \sin\vartheta)
  \mu}{D_m(\omega,\vartheta)},\\
  \bar{Q}_{m,\vartheta} &=
  \frac{(a_{m} - c_m \cos\vartheta - s_m  \sin\vartheta)
  \mu}{D_m(-\omega,\vartheta)},
\end{align}
\end{subequations}
where
\begin{eqnarray}
D_m(\pm\omega,\vartheta) = -\Omega \cos\vartheta +  k_m \sin\vartheta 
+\lambda \mp \omega\eta + (1-\alpha)\tilde\mu_\vartheta,
\end{eqnarray}
and
\begin{subequations}
\label{abc:def}
\begin{align}
a_{m} &= \int
        (Q_{m,\vartheta^{\prime}}-\alpha
        \bar{Q}_{m,\vartheta^{\prime}}) g(\vartheta') d\vartheta^{\prime},
\\
c_m  &= \int
       (Q_{m,\vartheta^{\prime}}-\alpha
       \bar{Q}_{m,\vartheta^{\prime}}) \cos\vartheta^{\prime}
       g(\vartheta') d\vartheta^{\prime},
\\
s_m  &= \int
       (Q_{m,\vartheta^{\prime}}-\alpha
       \bar{Q}_{m,\vartheta^{\prime}}) \sin\vartheta^{\prime}
       g(\vartheta') d\vartheta^{\prime}.
\end{align}
\end{subequations}
Substituting Eq.~\eqref{q:sol} in Eq.~\eqref{abc:def} we obtain a
characteristic equation for $(a_m,c_m,s_m)$:
\begin{eqnarray}
\begin{pmatrix}
I_m[1]-1 & -I_m[\cos\vartheta] & -I_m[\sin\vartheta] \cr
I_m[\cos\vartheta]  & -I_m[\cos^{2}\vartheta]-1  & -I_m[\cos\vartheta \sin\vartheta] \cr
I_m[\sin\vartheta] & -I_m[\cos\vartheta \sin\vartheta] & -I_m[\sin^{2}\vartheta] - 1 
\end{pmatrix}
\begin{pmatrix}
a_{m} \cr
c_m  \cr
s_m 
\end{pmatrix} = 0,
\label{stab:con}
\end{eqnarray}
where 
\begin{eqnarray}
I_m[f(\vartheta)]
=\int
f(\vartheta) g(\vartheta) \left[\frac{\mu}{D_m(\omega,\vartheta)}
  -\frac{\alpha\mu}{D_m(-\omega,\vartheta)} \right] d\vartheta
\end{eqnarray}
for arbitrary funciton $f(\vartheta)$.
Eq.~\eqref{stab:con} holds only when
\begin{align}
\mathrm{det}\begin{vmatrix}
I_m[1]-1 & -I_m[\cos\vartheta] & -I_m[\sin\vartheta] \cr
I_m[\cos\vartheta]  & -I_m[\cos^{2}\vartheta]-1  & -I_m[\cos\vartheta \sin\vartheta] \cr
I_m[\sin\vartheta] & -I_m[\cos\vartheta \sin\vartheta] & -I_m[\sin^{2}\vartheta] - 1 
\end{vmatrix} = 0.
\label{eq:det}
\end{align}
For given $m$, $\lambda$ and $\mu$ one can find a set of
$\Omega_m^\p{i}(\lambda,\mu)$ ($i=1,2,\ldots$) which satisfy
Eq.~\eqref{eq:det} and which are the frequencies
of the corresponding normal modes of collective neutrino oscillations. When
\begin{align}
\kappa_m^\p{i} = \mathrm{Im}(\Omega_m^\p{i})
\end{align} 
is positive, the corresponding normal mode is unstable and its
amplitude grows exponentially. If there exist multiple unstable
modes, the mode with the largest exponential growth rate,
\begin{align}
\kmax_m = \max(\kappa_m^\p{i}),
\label{eq:kmax}
\end{align}
will eventually dominate.

\section{Results in the linear regime}

\subsection{Numerical computation}
\begin{figure*}
$\begin{array}{cc}
\includegraphics[width=0.49\textwidth]{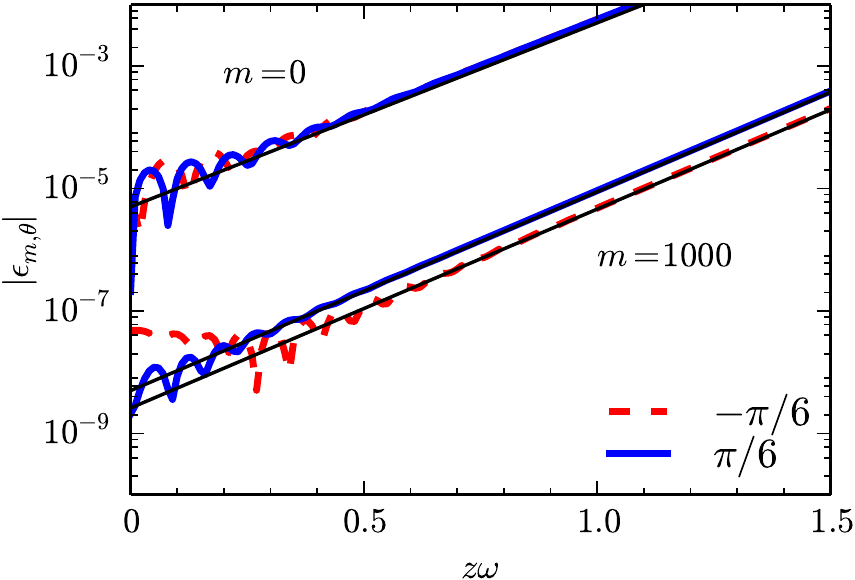} &
\includegraphics[width=0.49\textwidth]{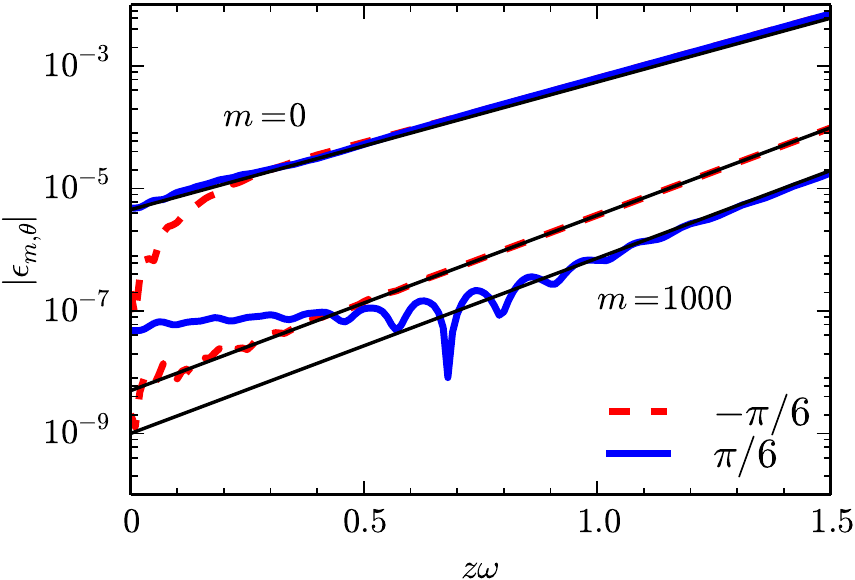}
\end{array}$
\caption{The evolution of $|\epsilon_{m,\vartheta}|$, the amplitudes of
  the off-diagonal elements of the neutrino moment 
  matrices $\rho_{m,\vartheta}(z)$, in terms of propagation distance
  $z$ for the inverted
  (left) and normal (right) neutrino mass hierarchies. The thick
  curves represent the numerical 
  solution to Eq.~\eqref{eq:moments} with $0$th and $1000$th moments
  only. The thin solid lines represent the exponential growth
  functions $\sim \exp(\kmax_m z)$ predicted by the linear stability
  analysis.  In these calculations we used the parameters 
  listed in Eq.~\eqref{eq:par}, and we took the matter
  potential $\lambda=0$ and neutrino potential $\mu/\omega=1500$ (left)
  and $3000$ (right) which is measured in the vacuum neutrino
  oscillation frequency  $\omega$.} 
\label{del:rho}
\end{figure*}

We develop a computer code to solve Eq.~\eqref{eq:moments}
numerically. In this code the continuous range of $\vartheta$ is
represented as $N$ discrete angle bins with central value
$\vartheta_i$ ($i=1,\ldots,N$) and equal interval $\Delta\vartheta$. 
For an arbitrary function $f(\vartheta)$ one has
\begin{align}
\int f(\vartheta) d\vartheta 
\longrightarrow \Delta\vartheta \sum_{i=1}^N f(\vartheta_i).
\end{align}

In this work we focus on the neutrino oscillations in the linear regime and
the cases with a simple angular distribution which has
isotropic neutrino fluxes within range $[-\thmax,\thmax]$, i.e.
\begin{align}
g(\vartheta) = \left\{\begin{array}{ll}
\frac{1}{2}\thmax^{-1} & \text{ if } \vartheta\in[-\thmax,\thmax],\\
0 & \text{ otherwise.}
\end{array}\right.
\label{eq:g}
\end{align}
We choose to present our results with the
following parameters
\begin{align}
\thmax=\pi/6, \quad
\alpha=0.8
\quad 
\text{and}\quad L=40\pi \omega^{-1}.
\label{eq:par}
\end{align}  
Because the evolution of different
neutrino moments is decoupled in the linear regime, it is
sufficient to include only the 0th and $m$th moments in studying
the evolution of the $m$th moment in this regime. [The 0th moment is
needed because it has large diagonal elements even in the linear regime. See
Eq.~\eqref{den:mat}.]

\begin{figure*}
$\begin{array}{c}
\includegraphics[width=.99\textwidth]{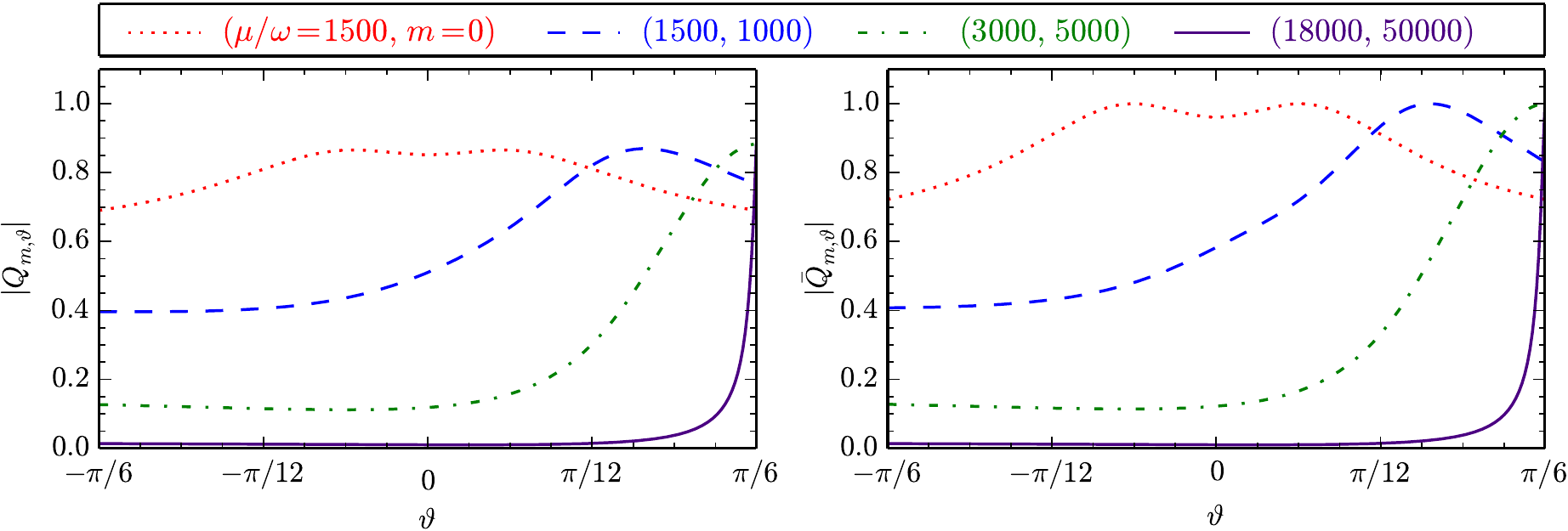}\\
\includegraphics[width=.99\textwidth]{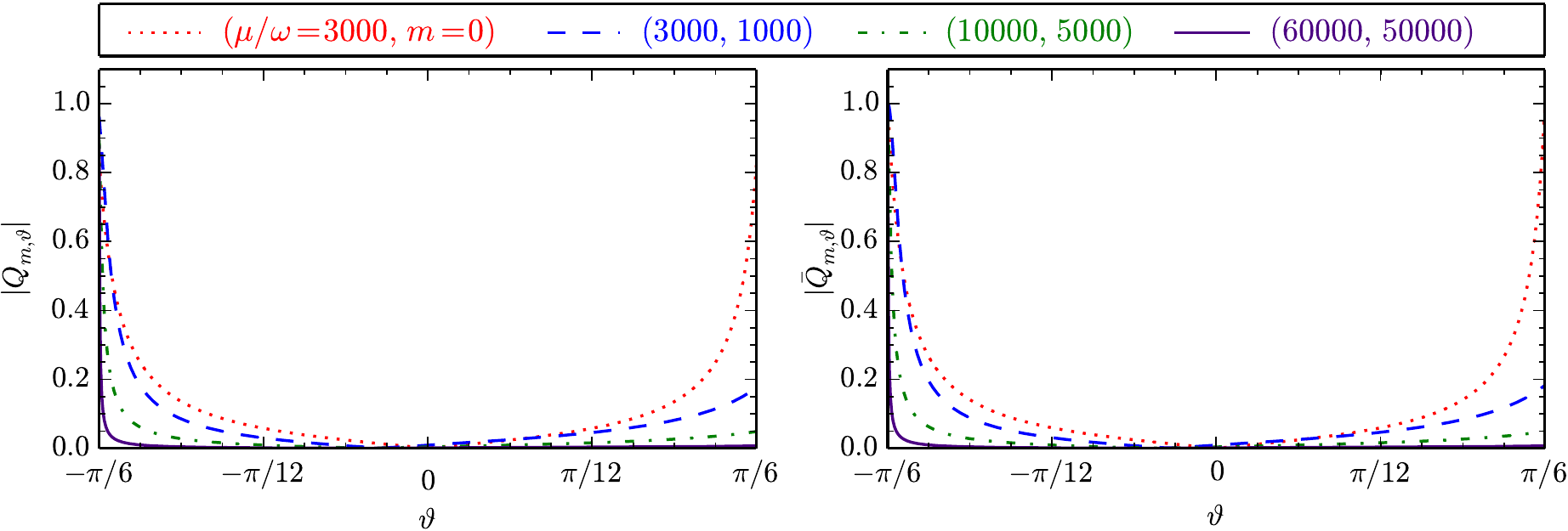}
\end{array}$
\caption{The amplitudes
  of the unstable modes of the $m$th neutrino moments 
  (in arbitrary scale) as functions of
  neutrino emission angle $\vartheta$
  which have the largest
  exponential growth rates in the linear regime
  at given neutrino
  number densities (indicated by $\mu=\sqrt2 G_F n_\nu$ which is
  measured in the vacuum neutrino oscillation frequency $\omega$). The top
  and bottom 
  panels are for the inverted and normal neutrino mass hierarchies,
  respectively. In these calculations we used the parameters
  listed in Eq.~\eqref{eq:par}, and we took the matter
  potential $\lambda=0$.}
\label{fig:q}
\end{figure*}

In Fig.~\ref{del:rho} we show the numerical solutions to
Eq.~\eqref{eq:moments} in two calculations with all but the 0th
and $1000$th moments being zero. In both calculations,
$|\epsilon_{m,\vartheta}|$, the
amplitudes of the 
off-diagonal elements of $\rho_{m,\vartheta}$, grow exponentially
which is understood as flavor instabilities.
As comparison we plot in Fig.~\ref{del:rho} the exponential
growth functions $\sim\exp(\kmax_m z)$ predicted 
by the flavor stability analysis, and they
agree with the numerical results very well.
As a further confirmation, we have compared the shapes of
$|Q_{m,\vartheta}|$ and $|\bar{Q}_{m,\vartheta}|$ obtained from flavor
stability analysis (the dotted and dashed curves in Fig.~\ref{fig:q})
with those of $|\epsilon_{m,\vartheta}|$ and $|\bar\epsilon_{m,\vartheta}|$ in 
numerical calculations (not shown),
and they also have good agreement.
However, to achieve numerical convergence a large number of angle bins
may be needed for the following reason.
  
\begin{figure*}
\includegraphics[width=0.99\textwidth]{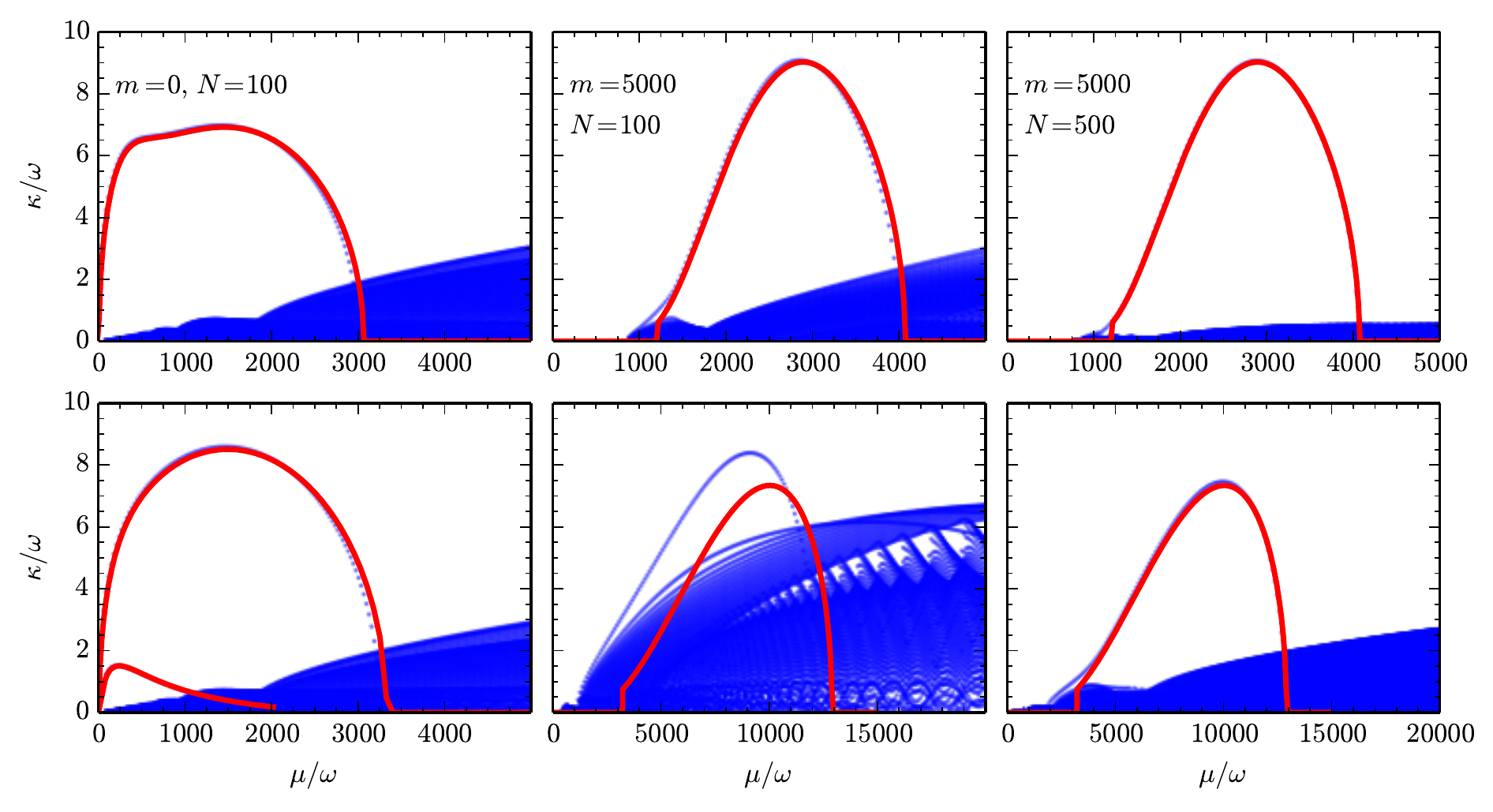}
\caption{The exponential growth rates $\kappa_m^\p{i}$ of the
  unstable collective modes of the $m$th neutrino moment as functions
  of neutrino self-interaction strength $\mu=\sqrt2 G_F n_\nu$
  in the discrete 
  angle-bin scheme with $N$ angle bins (as labeled and shown as the
  dotted curves) and in the continuum limit of 
  angular distribution  (solid curves), 
  respectively. Both $\kappa$ and $\mu$ are measured in the vacuum
  neutrino oscillation frequency $\omega$.  
  The top and bottom panels are for the inverted and
  normal neutrino mass hierarchies, respectively.
  In these calculations we used the parameters
  listed in Eq.~\eqref{eq:par}, and we took the matter
  potential $\lambda=0$.}
\label{spur:1}
\end{figure*}

As pointed out in
Ref.~\cite{Sarikas:2012ad}, there can exist many spurious flavor
instabilities in the numerical implementation using the discrete
(angle-bin) scheme. This can be 
seen from the discretized version of Eq.~\eqref{off:term}:
\begin{subequations}
\label{lin:eom}
\begin{align}
i\cos\vartheta_{i} \partial_{z} \epsilon_{m,\vartheta_{i}}  
&=  [k_m  \sin\vartheta_{i} +\lambda  - \omega\eta 
+(1-\alpha)\tilde\mu_{\vartheta_i}]
\epsilon_{m,\vartheta_{i}}  
\nonumber\\
&\quad
- \mu \Delta\vartheta \sum_{j} [1-\cos(\vartheta_i-\vartheta_j)]
(\epsilon_{m,\vartheta_{j}} -\alpha
  \bar{\epsilon}_{m,\vartheta_{j}} )
g(\vartheta_j),\\ 
i\cos\vartheta_{i} \partial_{z} \bar\epsilon_{m,\vartheta_{i}}  
&=  [k_m  \sin\vartheta_{i} +\lambda  + \omega\eta 
+(1-\alpha)\tilde\mu_{\vartheta_i}]
\bar\epsilon_{m,\vartheta_{i}}  
\nonumber\\
&\quad
- \mu \Delta\vartheta \sum_{j} [1-\cos(\vartheta_i-\vartheta_j)]
(\epsilon_{m,\vartheta_{j}} -\alpha
  \bar{\epsilon}_{m,\vartheta_{j}} )
g(\vartheta_j),
\end{align}
\end{subequations}
or
\begin{align}
i \partial_z \boldsymbol{\epsilon}_m = \mathsf{\Lambda}_m \cdot
  \boldsymbol{\epsilon}_m, 
\end{align}
where $ \boldsymbol{\epsilon}_m = 
(\epsilon_{m,\vartheta_{1}},
\bar{\epsilon}_{m,\vartheta_{1}},
\epsilon_{m,\vartheta_{2}},
\bar{\epsilon}_{m,\vartheta_{2}}, \ldots,
\epsilon_{m,\vartheta_{N}}, 
\bar{\epsilon}_{m,\vartheta_{N}})^T$ 
is a $2N$-dimensional vector,
and $\mathsf{\Lambda}_m$ is a
$2N\times 2N$ real matrix. Matrix $\mathsf{\Lambda}_m$
has $2 N$ eigenvalues $\Omega_m^\p{i}$ ($i=1,2,\ldots,2N$)
each of which corresponds to the collective oscillation frequency 
of a collective mode  in the discrete
scheme. Many of these 
collective modes can be unstable, i.e.\ with
$\kappa_m^\p{i}=\text{Im}(\Omega_m^\p{i})>0$. Only a few of the
unstable modes correspond to the physical instabilities in the continuum limit
(of the $\vartheta$ distribution), and the rest of them are
``spurious'' or the artifact of the numerical implementation.  

In Fig.~\ref{spur:1} we plot the exponential growth rates $\kappa^\p{i}_m$ of
all the unstable collective modes both in the discrete scheme and in
the continuum limit for the 0th and $5000$th moments, respectively. This
figure shows that spurious instabilities (in the discrete scheme) can
dominate the physical instabilities (in the continuum limit) on
small distance scales and/or large
neutrino number densities (i.e.\ large $|m|$ and/or $\mu$). In some
extreme cases, e.g., the bottom middle panel of 
Fig.~\ref{spur:1} where $\eta=+1$, $m=5000$ and $N=100$, none of the
collective modes in the discrete scheme matches the ones in the
continuum limit. This is likely due to the fact that  $Q_{m,\vartheta}$ and
$\bar{Q}_{m,\vartheta}$ become sharply peaked functions of
$\vartheta$ at large $|m|$ and/or $\mu$, which requires more angle
bins to resolve (see Fig.~\ref{fig:q}). Indeed, the comparison between
the middle and right 
panels of Fig.~\ref{spur:1} show that the spurious instabilities are
more suppressed when more angle bins are employed.

\subsection{Flavor instabilities and matter effect}
\begin{figure*}
$\begin{array}{cc}
\text{IH} & \text{NH}\\
\includegraphics[width=0.49\textwidth]{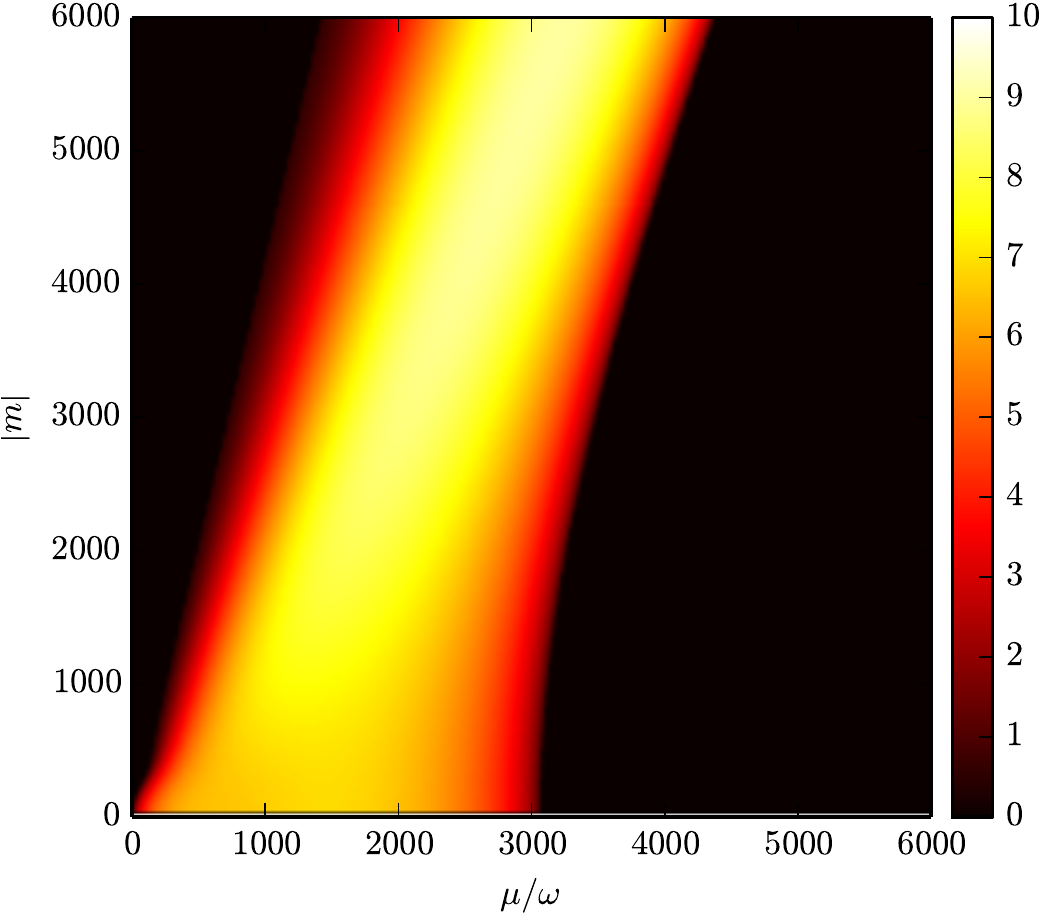}
&\includegraphics[width=0.49\textwidth]{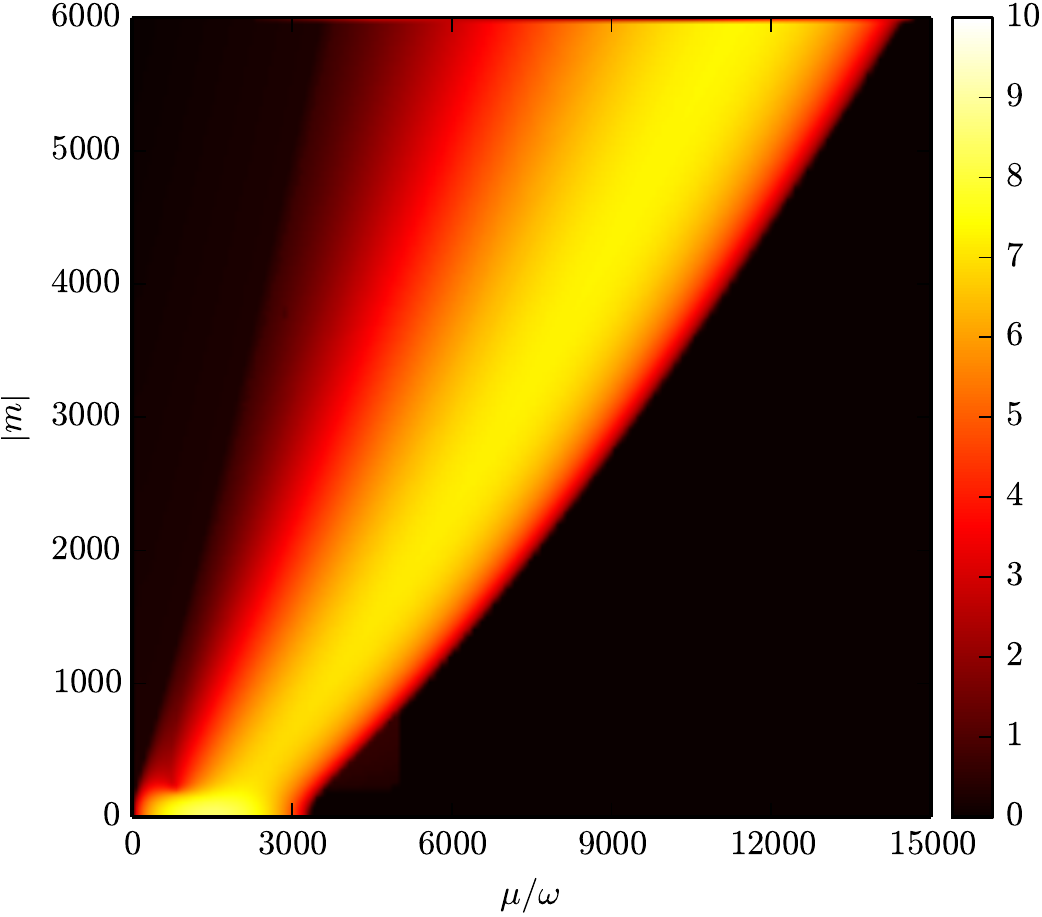} \\
\includegraphics[width=0.49\textwidth]{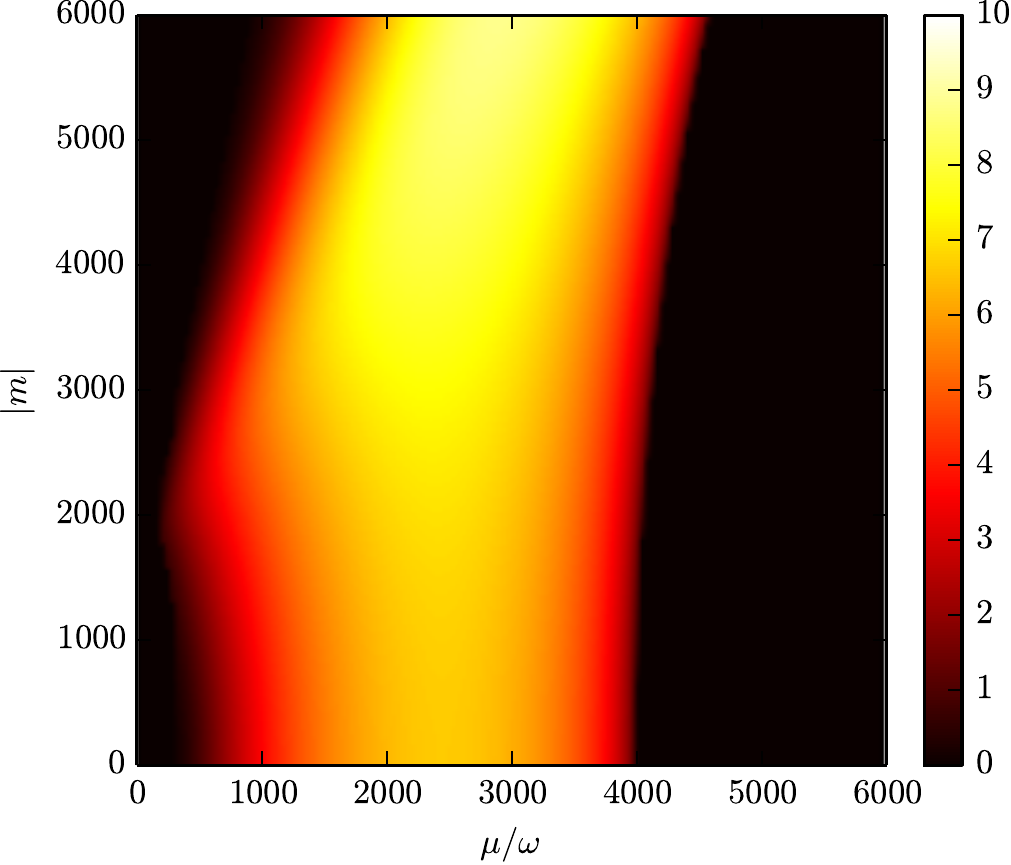}
&\includegraphics[width=0.49\textwidth]{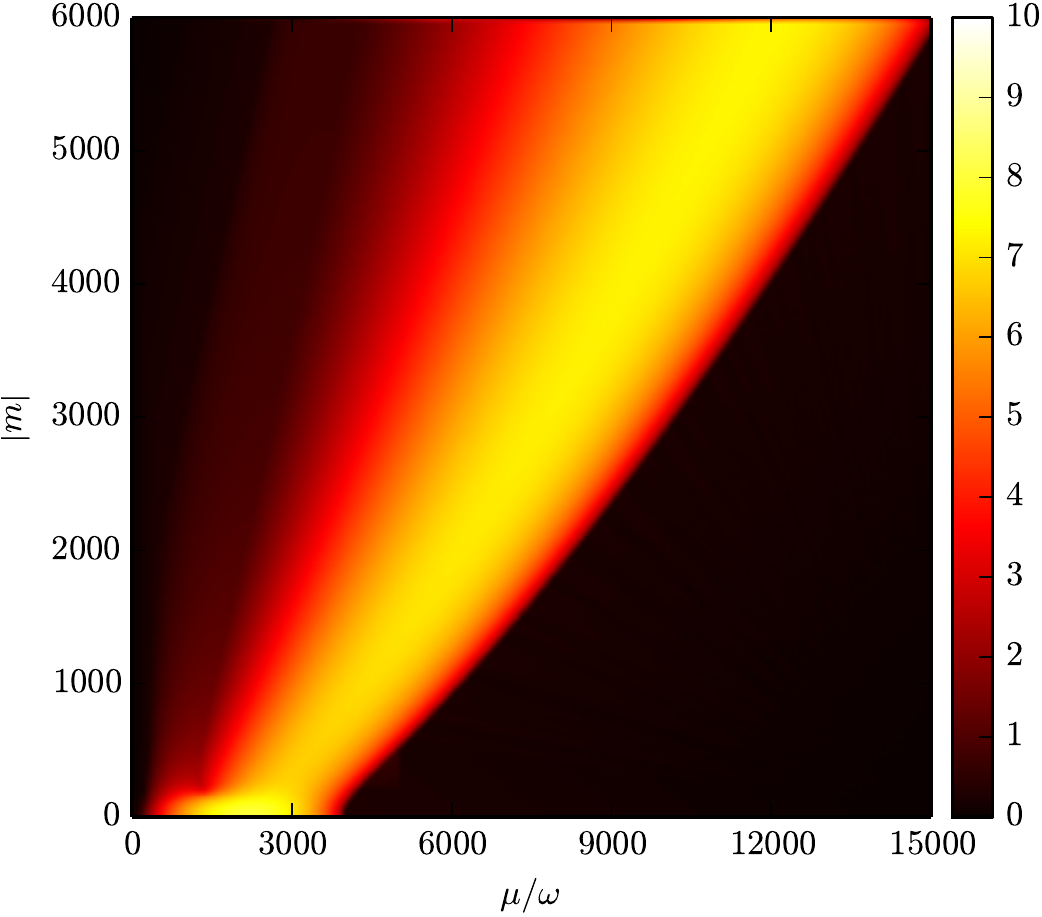}
\end{array}$
\caption{Maximum exponential growth rate
  $\kappa^\mathrm{max}_m(\lambda,\mu)$ (indicated by the color scale)
  of the neutrino collective oscillation modes in the multi-angle Line model
  as a function of moment
  index $m$ and the neutrino self-coupling strength
  $\mu=\sqrt{2}G_F n_\nu$. Both $\kappa$ and $\mu$ are measured in the
  vacuum neutrino oscillation frequency $\omega$.
  The left
  and right panels are for the inverted and normal
  neutrino mass hierarchies, respectively, and the top and bottom
  panels are for $\lambda=\sqrt2G_F n_e=0$ and $200\omega$,
  respectively.
  In these calculations we assume 
  isotropic neutrino fluxes within angular range $\vartheta\in[-\pi/6,\pi/6]$,
  and we used the parameters listed in Eq.~\eqref{eq:par}.}
\label{kappa:plot}
\end{figure*}

We have solved the flavor instabilities of the multi-angle Line model
using the angular distribution in Eq.~\eqref{eq:g} and the parameters listed in
Eq.~\eqref{eq:par}. The results for the neutrino gas in the absence of
matter are shown in the
upper panels of Fig.~\ref{kappa:plot}. 
From this figure one can see that, unlike the two-beam Line model
\cite{Duan:2014gfa}, the flavor instabilities in the multi-angle model
depends on the neutrino mass hierarchy, and collective
oscillations can begin at larger neutrino density in NH than in
IH. One also sees that both
$\mu_m^\text{max}$ and $\mu_m^\text{min}$, the maximum and
minimum $\mu$ values where the $m$th modes are unstable, seem to increase
linearly with $|m|$. In contrast, both
$\mu_m^\text{max}$ and $\mu_m^\text{min}$ increase linearly with
$\sqrt{|m|}$ in the two-beam model.%
\footnote{The definition of neutrino self-coupling strength $\mu$ in
  \cite{Duan:2014gfa} has taken into account of the geometric factor
  $1-\cos(\vartheta-\vartheta')$ and is equivalent to
  $\tilde\mu_\vartheta$ in this paper. For the angular distribution in
  Eq.~\eqref{eq:g} $\tilde\mu_0 = (1-\sin\thmax/\thmax)\mu \approx 0.045\mu$.  
}
This implies that, for sufficiently large $|m|$, flavor instabilities
can develop at even larger neutrino densities in the multi-angle model than
in the two-beam model.

Unlike in the two-beam model, the presence of matter can affect
collective oscillations in the multi-angle model because the neutrinos
propagate in different directions can travel through different distances
between two lines that are parallel to the neutrino Line.
In the lower panels of Fig.~\ref{kappa:plot} we show the flavor
instabilities in the multi-angle Line model with $\lambda =
200\omega$. Similar to the situation in the spherical neutrino Bulb
model for supernova \cite{Banerjee:2011fj,EstebanPretel:2008ni}, both
$\mu_m^\text{max}$ 
and $\mu_m^\text{min}$ of the homogeneous mode
(i.e.\ with $m=0$) shift to larger values in the presence of a large
matter density in both NH and IH. However, $\mu_m^\text{min}$ of
inhomogeneous modes actually shifts to
smaller values for both NH and IH when $|m|$ is sufficiently large.

\section{Discussion}

We have used both the numerical method and the linear stability
analysis to investigate collective neutrino oscillations in the
multi-angle Line model in the linear regime where the neutrino flavor
transformation is still small. Although the Line model does not
represent any real physical environment, the study of this toy model
can provide insights into the important differences between the models of one
spatial dimension (e.g.\ the neutrino Bulb model for supernova) and
multi-dimension models.

An important goal of this work is to check if the inhomogeneous
collective modes are suppressed in the multi-angle environment because
of the high neutrino densities which is known to exist
in the Bulb model \cite{Duan:2010bf,EstebanPretel:2008ni}. 
Somewhat surprisingly, our work suggests
that, in the absence of ordinary matter, 
inhomogeneous collective modes on small scales are not only not
suppressed in the multi-angle environment, but can become
unstable at larger neutrino densities than in the two-beam model.

We also examined whether the presence of a large matter density can
suppress collective oscillations in the two-dimensional Line model as
in the one-dimensional Bulb model \cite{EstebanPretel:2008ni}. 
Our study shows that the presence of ambient matter does suppress
inhomogeneous oscillation modes on large distance scales in the Line
model as it occurs to the homogeneous modes in the Bulb
model. However, it appears that the inhomogeneous modes on very small
scales can occur at smaller neutrino number densities with
ambient matter than without. In addition, the flavor unstable region
of the certain inhomogeneous modes can extend to the regime of lower
neutrino densities than that for the homogeneous mode.

We have shown that, as in the Bulb model, there exist spurious
oscillations in the numerical implementation of the multi-angle Line model
if the discrete angle-bin scheme is employed. The problem of 
spurious oscillations appear to be more severe at higher neutrino
densities and on smaller distance scales. Although this problem 
can be mitigated by using more angle bins, it does add complications to
the already challenging task of computing collective neutrino oscillations
near astrophysical neutrino sources such as core-collapse supernovae
and black-hole accretion discs.  It is probably helpful to develop the 
multipole expansion method similar to that for the Bulb model
\cite{Duan:2014mfa}.

Our work has focused on the neutrino flavor instabilities in the linear
regime. However, not every flavor 
instability in the linear regime can result in significant neutrino
flavor transformation. For example, in the realistic supernova environment, the
neutrino density decreases as neutrinos travel away from the center of the
supernova which results in the shift of the instability region. It is,
therefore, possible that a collective oscillation mode does not grow
all the way to the nonlinear regime during the finite distance
interval where it is unstable. We have considered the mixing of two
neutrino flavors only, which can be quite different from the neutrino
flavor transformation of three flavors \cite{Friedland:2010sc}. Ultimately,
the phenomenon of collective neutrino oscillations has to be studied
in realistic, multi-dimensional models for compact objects
such as core-collapse supernovae and black-hole accretion discs before
one can fully understand the impact of neutrino oscillations to these
extreme environments. 

\acknowledgments{
We thank V.~Noormofidi and L.~Ma for useful
discussions. We also appreciate the
hospitality of INT/UW where part of this work was done.
This work was supported by DOE EPSCoR grant \#DE-SC0008142 at UNM.}
\bibliography{line-ma}
\end{document}